\documentclass{PoS}

\title{Nonlinearly Realized Gauge Theories\\ for LHC Physics}

\ShortTitle{Nonlinearly Realized Gauge Theories for LHC Physics}

\author{Daniele Binosi\\
        European Centre for Theoretical Studies in Nuclear Physics
and Related Areas (ECT*) and Fondazione Bruno Kessler, Trento, Italy\\
       E-mail: \email{binosi@ect.it}
}

\author{Daniele Bettinelli,
\speaker{Andrea Quadri}\\
        Univ. di Milano and INFN, Sez. di Milano, Italy\\
        E-mail: \email{andrea.quadri@mi.infn.it}\\
        }

\abstract{We consider a minimal nonlinearly realized electroweak theory where
mass generation happens {\em {\`a la}} St\"uckelberg. Deformation
of the nonlinearly realized gauge symmetry is controlled by functional
methods. The Weak Power Counting allows to select uniquely the Hopf algebra
of the theory and gives definite predictions on the Beyond-the-Standard Model (BSM) sector of the theory: the latter includes one CP-odd and two charged physical scalars (in addition to the Higgs-like CP-even resonance). The model
interpolates between a purely St\"uckelberg and a Higgs scenario. It can be used in order to check whether the presence of 
a St\"uckelberg mass component can already be excluded on the basis of 
the existing LHC7-8 data.}

\FullConference{The European Physical Society Conference on High Energy Physics \\
                 18-24 July, 2013\\
                 Stockholm, Sweden}

\begin{document}


The LHC discovery of a new physical scalar resonance
of mass about $125$ GeV has opened the way to
the experimental verification of the 
electroweak Spontaneous Symmetry Breaking (SSB) mechanism
realized in Nature.

The favourite candidate after the LHC7-8 data is the 
Standard Model (SM)
Higgs mechanism. It is therefore timely to check
whether one can already exclude,  on the basis
of the current LHC data, one of the most 
ancient competitors to the SSB {\em \`a la} Higgs,
namely the St\"uckelberg mechanism~\cite{Stuck}.
The latter allows to introduce gauge-invariant mass terms
for the gauge bosons and matter fermions by using an operatorial gauge
transformation controlled by 
a group element $\Omega \in \mathrm{SU(2)}$, parameterized by
the independent coordinates $\phi_a$,
in a way preserving physical unitarity~\cite{Ferrari:2004pd}.
The gauge symmetry is nonlinearly realized, as a consequence
of the nonlinear SU(2) constraint.

        Nonlinearity is reflected into the severe UV divergences
        of the theory. Since the discovery of the Local Functional
        Equation (LFE)~\cite{LFE}, it has been known that 1-PI amplitudes
        involving at least one $\phi_a$-leg (descendant amplitudes) 
        are uniquely fixed in terms of amplitudes with
        no external $\phi_a$-legs (ancestor amplitudes).
        The LFE controls the deformation of the classical
        nonlinearly realized gauge symmetry, induced by radiative
        corrections.

        Moreover, while there are infinitely many divergent
        descendant amplitudes already at one loop level,
        the number of divergent ancestor amplitudes is finite
        at each loop order, provided that the Weak Power Counting (WPC)
        holds~\cite{WPC}. 
        
        The WPC selects uniquely the Hopf algebra of the theory.
        The number of UV divergent ancestor amplitudes increases
        order by order in the loop number and thus the model
        is not renormalizable by power-counting.
        The implementation of the program based on the LFE and 
        the WPC has been developed in Yang-Mills theory
        for SU(2) in~\cite{NLtheoryYM}
        and for the semisimple group SU(2)$\times$U(1) in~\cite{NLtheoryEW}. 
        The electroweak model 
        with physical scalar resonances has been studied in~\cite{Binosi:2012cz}.
         
        Since these theories are not power-counting renormalizable,
        their UV-completion (if any) is not known.
        Therefore they can be used as an
        effective low-energy description, valid up to a certain
        energy scale $Q$. Loosely speaking, one expects that 
        this is the scale where violation of unitarity occurs
        in physical scattering amplitudes.

        Interestingly enough, in the nonlinearly realized electroweak 
        gauge theory in the presence of physical scalar resonances~\cite{Binosi:2012cz}, $Q$ can be pushed at arbitrarily high energy,
        if the parameters of the theory are fine-tuned towards the SM-like
        region.
        
        The inclusion of a single physical scalar resonance
        in the nonlinearly realized electroweak model is not possible
        if the WPC is to hold.
        The minimal choice is a SU(2) doublet of fields~\cite{Binosi:2012cz}
        $ \chi= \chi_0 + i \chi_a \tau_a \, . $
        Thus there is also a mass invariant generated as in the usual Higgs mechanism
        from the SU(2) doublet of scalars $\chi$:
        $$
        {\cal L}_{mass,linear} = \frac{1}{4} ~{\rm Tr}  (D_\mu \chi)^\dagger D^\mu \chi \, .
        $$
        $\chi_0$ acquires a vacuum expectation value $v$, so that 
        it is split according to $\chi_0  = v + X_0$.
        SSB, triggered by a suitable
        quartic potential, must occur for the SU(2) doublet along
        the $\chi_0$-component. The reason is that otherwise
        one cannot accommodate for the suppression of the
        decay width of $X_0 \rightarrow \gamma\gamma$ with respect to
        the decay modes $X_0 \rightarrow VV$, $V=W,Z$ (which,
        without SSB, would be radiatively generated as well).
        The masses of the $W$ and $Z$ bosons are thus given by
        $$
        M_W = \frac{gv}{2} \sqrt{1 + A \frac{f^2}{v^2}} \, , ~~~~
        M_Z = \frac{Gv}{2}  \sqrt{1 + \frac{f^2}{v^2} \Big ( A + \frac{B f^2}{2} \Big )} \, .
        $$
        $g,g'$ are the SU(2) and hypercharge U(1) coupling constants respectively and
$G=\sqrt{g^2+{g'}^2}$.

        The parameters $A,B$ describe the contribution to the gauge boson masses 
        induced by the St\"uckelberg mechanism.
        In particular $B$ controls the violation of custodial symmetry in the gauge boson sector. Such a violation is a characteristic feature of nonlinearly realized theories
based on the WPC~\cite{Quadri:2010uk}.
        If $A=0$, $B=0$ one gets back the SM scenario where the 
electroweak SSB is realized
through the linear Higgs mechanism. In this case the $\phi_a$
decouple and the Goldstone bosons are to be identified with the
$\chi_a$ fields.
At $B=0$ and $A \neq 0$ 
a fraction of the mass of the gauge bosons is generated via
the St\"uckelberg mechanism while still preserving the Weinberg relation.
Finally, $A \neq 0$ and $B \neq 0$ corresponds to the most
general St\"uckelberg case with two independent mass terms
for the $W$ and the $Z$ bosons. 

 The theory 
       makes definite predictions on the field content of the BSM sector: 
       there must be four scalar resonances, two charged
       ones and two neutral ones, one CP-even (to
       be eventually identified with the $125$ GeV
       resonance recently discovered at LHC) and
       one CP-odd.

%
%
%
       Current LHC data clearly favour a scenario where new physics
       contributions, resulting in deviations from the SM values,
       are small~\cite{fit}. 
      The physically interesting case is therefore
       achieved in the small $A$ limit at $B=0$~\cite{Bettinelli:2013hia}.
We remark that if even a fraction of the gauge boson masses is generated
        by the St\"uckelberg mechanism, tree-level unitarity
        in the scattering of longitudinally polarized
        W bosons
         is
        violated, despite the exchange of a physical scalar resonance.
       This feature is a characteristic footprint of
       nonlinearly realized theories.

\end{document}